\documentclass[useAMS,usenatbib,psfig]{mn2e}
\usepackage{graphicx}
\usepackage{psfig}
\usepackage{epsf}

\def\H0{{\rm ~km~s^{-1}~Mpc^{-1}}}

\def\la{\mathrel{\hbox{\rlap{\hbox{\lower4pt\hbox{$\sim$}}}{\raise2pt\hbox{$<$}}}}}
\def\ga{\mathrel{\hbox{\rlap{\hbox{\lower4pt\hbox{$\sim$}}}{\raise2pt\hbox{$>$}}}}}
\def\d25{D$_{25}$}

\title[The ultimate outcome of BH--NS mergers]
{The ultimate outcome of black hole -- neutron star mergers}
\author[M.B.~Davies, A.J.~Levan, A.R.~King]{
Melvyn B.~Davies$^1$, Andrew J.~Levan$^2$, Andrew R.~King$^2$\\
$^1$Lund Observatory, Box 43, SE--221 00 Lund, Sweden.\\
$^2$Department of Physics and Astronomy, University of Leicester,
Leicester, LE1~7RH, UK.\\
}

\date{Submitted 18 August 2004}

\pagerange{\pageref{firstpage}--\pageref{lastpage}}
\pubyear{2004}

\begin{document}

\maketitle

\label{firstpage}

\begin{abstract}
We present a simple, semi--analytical description for the final stages of 
mergers of black hole (BH) -- neutron star (NS) systems. Such systems are of
much interest as gravitational wave sources and gamma--ray burst
progenitors. Numerical studies show that in general the neutron star is not
disrupted at the first phase of mass transfer. Instead, what remains
of the neutron star is left on a wider, eccentric, orbit. 
We consider the evolution of such systems as they lose angular
momentum via gravitational radiation and come into contact for
further phases of mass transfer.  During each mass transfer event the neutron
star 
mass is reduced until a critical value where mass loss leads to
a rapid increase in the stellar radius. At this point Roche lobe overflow
shreds what remains of the neutron star,
most of the mass forming a disc around the black hole.
Such a disc may be massive enough to power a gamma--ray burst.
The mass of the neutron star at the time of disruption (and
therefore the disc mass) is largely
independent of the  initial masses
of the black hole and neutron star, indicating that BH--NS star mergers may
be standard candles.
\end{abstract}


\section{Introduction}
Binary systems consisting of a pair of compact objects (neutron
stars (NS) or black holes (BH)) have long been of interest as tests
of relativity and gravitational wave sources (Hulse and Taylor, 1975)
or gamma-ray burst (GRB) progenitors (see e.g. Fryer, Woosley
\& Hartmann, 1999).
Their formation requires that the system remains bound through the
two supernovae creating the compact objects. While there exist many
possible channels for the formation of NS--NS binaries, relatively
few exist for the formation of BH--NS (or BH--BH) systems
(Belczynski et al. 2002). The basic mechanism for the formation
of BH--NS systems is non--conservative mass transfer between the
two massive stars in a binary, common envelope evolution (either 
before or after the first SN event) to dissipate angular momentum and 
mass, and then finally a second SN. These systems subsequently
lose further angular momentum via gravitational radiation and
a substantial fraction merge within  Hubble time 
(See Portegies Zwart \& Yungelson 1998; Belczynski et al. 2002 for
more information on the possible channels for BH--NS creation).
While it is now thought that at least the majority of GRBs
with durations longer than 2s result from core collapse of
massive stars to supernovae (Hjorth et al. 2003; Stanek et al. 2003)
binary mergers remain a likely candidate for the distinct class
of GRBs with durations less than 2s (Kouveliotou et al. 1993).
If this is the case then BH--NS mergers may contribute significantly
to the short--GRB population.

High--resolution hydrodynamical simulations have proved a valuable
tool in probing the last few orbits and final mergers of compact
binary systems (Janka et al. 1999; Rosswog et al. 1999,  
Rosswog \& Davies 2002; Rosswog \& Liebendorfer 2003)
and as their potential as short-GRB progenitors (e.g. Rosswog \&
Ramirez-Ruiz 2003). Much effort has gone into the understanding
of double neutron star systems as progenitors, while BH--NS systems
have also received attention (Kluzniak \& Lee
1998; Portegies Zwart 1998;
 Lee 2000,2001; Janka et al. 1999; Rosswog, Speith \& Wynn 2004). 
To be a viable GRB progenitor a 
compact object merger must make 
a massive torus ($\sim$ 0.1 - 1 M$_{\odot}$) surrounding a BH, or
supramassive neutron star (e.g. Ruffert \& Janka 1999; Veitra \& Stella 1998). 
The subsequent extraction of energy from this
torus provides power for the GRB, possibly via neutrino--antineutrino 
annihilation.  The formation of this torus is vital to the production of a GRB.
In collapsar models the formation of a massive torus is straightforward since,
in contrast to the mergers, a large mass (0.1 - 5 M$_{\odot}$)
reservoir is available from the
progenitor star (MacFadyen,  Woosley \& Heger, 2001). 
In NS--NS systems of approximately
equal mass it is also possible to form a massive disk
(see e.g. Rosswog \& Liebendorfer 2003). However
in the case of BH--NS systems the large mass ratio can dramatically
affect the outcome of the merger. For soft equations of state (e.g. 
polytropes with $\Gamma=2$) it is possible to have complete 
disruption on the first approach and thus is a reasonable
mechanism for GRB production (Lee \& Kluzniak 1999). However, for more realistic
 (harder) equations of state, such as that of Shen et al. (1998), 
the situation is less clear. 
When the neutron star spirals into contact with the black hole 
for the first time, 
a short burst of mass transfer on to the black hole
 follows, with mass transfer rates
in excess of 100 M$_{\odot}$ s$^{-1}$. Crucially, much of the angular
momentum of the transferred gas is fed back into what remains of the neutron
star, which is then left on an eccentric, and wider, orbit. This orbit
may be sufficiently wide that the neutron star no longer fills  its Roche 
lobe. The system then spirals together as angular momentum is lost
 via the emission of gravitational radiation. 
Mass transfer begins again once the system comes into contact.
The ultimate
fate of the neutron star remains an open question. It may disrupt either at
the onset of the second phase of mass transfer, 
(eg Ruffert \& Janka 2003),
or once mass transfer has reduced the NS 
to the minimum mass allowed by the NS equation of state
(eg Kluzniak \& Lee 1998, Colpi \& Wasserman 2002).

In this paper, we use a semi--analytical approach to model 
episodes of mass transfer from the NS to the BH.
Each episode results in the transfer of mass to the BH
but with some fraction of that material's angular momentum being
fed back into the NS, which is left in a wider, eccentric orbit.
The subsequent trajectories of the two stars are then calculated
allowing for the effects of gravitational radiation until mass
transfer occurs again.  The process is repeated until 
the supply of material is exhausted.

In section 2, we calculate the time required for the system 
to come into contact again after a burst of mass transfer. In section 3,
we compute the evolution of the system through a number of phases
of mass transfer. We consider the likely fate of the system once
the neutron star reaches its minumum mass in section 4. We draw
our conclusions in section 5.

\section{Evolution into later contact via gravitational radiation}

Often, after the first phase of mass transfer, what remains
of the neutron star is on an eccentric orbit
with a semi--major axis which
is sufficiently large that the neutron star no longer fills its Roche
lobe (see eg Lee 2000). Angular momentum and energy loss via 
gravitational radiation  brings
the system back into contact for a second phase of mass transfer.
The time required to bring the system back into contact is
a function of the eccentricity and semi--major axis of the orbit,
and the masses of the black hole and neutron star. The subsequent
evolution of the binary separation $a$ and eccentricity $e$ can
be calculated using the following expressions (Peters 1964).

\begin{eqnarray}
{da \over dt} & = & - { 64 G^3 M_{\rm bh} M_{\rm ns} (M_{\rm bh} + M_{\rm
ns})  
\over 5 c^5 a^3 (1 - e^2)^{7/2}}
\times \nonumber \\
&& \left( 1 + {73 \over 24} e^2 + {37 \over 96} e^4 \right)
\end{eqnarray}

\begin{equation}
{de \over dt} = - {304 G^3 M_{\rm bh} M_{\rm ns} (M_{\rm bh} + M_{\rm ns})
e \over 15 c^5 a^4 (1 - e^2)^{5/2}}
\times \left( 1 + {121 \over 304} e^2 \right)
\end{equation}

Taking typical values of $a = 150$ km, and $e = 0.2$, the timescale
for the system to return to contact ranges from about 0.025 s
(for $M_{\rm bh} = 14 M_\odot$) to 1 s (for $M_{\rm bh} = 2 M_\odot$).
When the system comes into contact again, another pulse of mass transfer
occurs. We note that the system is spiralling together
rapidly when mass transfer begins, with typical values for
$\delta a/a \simeq 0.05 - 0.10$ in one orbital period; the 
subseqent mass transfer is therefore dissimilar from that encountered
in the stable evolution of circular binaries. A large fraction
of the material from the neutron star is stripped, and what
remains gets a kick, leaving it (again) on a somewhat
eccentric orbit with an increased semi--major axis.

\section{Evolution via mass transfer down to the minimum mass.}

In this section we model the evolution of a BH--NS binary
using a two--body code, treating gravitational
radiation in the quadrupole approximation, and 
allowing for mass transfer when the neutron star fills its
Roche lobe, assuming the mass is transferred instantly to the 
black hole, but (crucially) returning some fraction of the
angular momentum of the transferred material back to the
neutron star on the timescale of the  orbital period when
mass transfer occurs.

The energy released in gravitational radiation is estimated by
(see eg Zhuge et al 1994)

\begin{equation}
L_{\rm gw} = {G \over 5 c^5} I_{\rm ij}^{(3)} I_{\rm ij}^{(3)}
\end{equation}
where $I_{\rm ij}^{(3)}$ is the third time derivative of the 
reduced quadrupole moment, $I_{\rm ij}$
and we sum over repeated indices. As the BH--NS binary
can be defined to sit in the xy--plane,  we need 
only  consider three components of  $I_{\rm ij}$, namely

\begin{equation}
I_{\rm xx} = \sum_{p=1}^2 M_p \left( x_p^2 - r_p^2 /3 \right)
\end{equation}

\begin{equation}
I_{\rm yy} = \sum_{p=1}^2 M_p \left( y_p^2 - r_p^2 /3 \right)
\end{equation}

\begin{equation}
I_{\rm xy} = \sum_{p=1}^2 M_p x_p y_p
\end{equation}

By computing the third time derivatives of the quadrupole moments,
we can calculate the rate of energy loss via gravitational radiation.
This energy loss rate can be applied to the two bodies via a drag force
(see eg Zhuge et al 1994).

Mass transfer occurs when the neutron star fills its Roche lobe,
ie when the separation is given by 

\begin{equation}
a_{\rm contact} = {R_{\rm ns} \over 0.49} \left( { M_{\rm ns}
+ M_{\rm bh} \over M_{\rm ns} } \right)^{1/3}
\end{equation}
where we make the approximation that the neutron--star
radius $R_{\rm ns} =$ 15 km for all neutron star masses. As we will
see shortly, the rapid increase in the actual radius for a real 
neutron star once its mass falls below about $0.2$ M$_\odot$
plays a crucial role in the subsequent evolution of the system.

To calculate the mass transferred from the neutron star,
$\delta M_{\rm ns}$, we estimate the change in separation 
occuring over half an orbit, i.e.

\begin{equation}
\delta a =  {1 \over 2} \left( \dot{a} \times \tau_{\rm orb} \right)
\end{equation}
and then assume that all material outside the Roche lobe at
a separation $a_{\rm contact}  - \delta a$ is transferred.
Given that the equation of state for the gas in a neutron star
is very hard (i.e. the gas is very resistant to compression), we
approximate a neutron star as a sphere of constant density.
Hence the mass transferred is given by

\begin{equation}
\delta M_{\rm ns} = M_{\rm ns} { \left( a_{\rm contact}^3
- (a_{\rm contact}  - \delta a)^3 \right) \over
a_{\rm contact}^3 }
\end{equation}

We make the approximation that the mass transfer is instantaneous, but return
half of the material's angular momentum to the neutron star
over the timescale of one orbital period. 
This injection of angular momentum leaves
the neutron star on a somewhat eccentric and, crucially, somewhat
wider orbit. Typically, we find that the neutron star no longer
fills its Roche lobe. We assume that no mass transfer occurs until
the system has spiralled in to contact when we repeat the procedure
described above. Thus the material within the neutron star
is transferred to the black hole via discrete bursts of mass transfer
separated by relatively long quiescent periods whilst 
the neutron star spirals in to contact via the emission of gravitational
radiation. In reality, there may well  be some mass loss from the 
neutron star during the quiescent periods. However, computer simulations 
show that this rate of mass loss is relatively small
(eg Rosswog et al 2004). Such low rates of mass loss may be due to oscillations
within the neutron star induced by the more violent 
phases of mass transfer occuring when the neutron star fills its
Roche lobe.

We show the evolution of one BH--NS binary in Figs 1 and 2.
Here we have taken the black hole mass, $M_{\rm bh} = 14 M_\odot$
and $M_{\rm ns} = 1.4 M_\odot$.
We begin the integration of the orbit with a separation of 300 km.
In Fig. 1 we show the separation as a function of time around
the onset of the first contact. Over half of the material
of the neutron star is transferred to the black hole
at a time of $\sim 0.56$ s. What remains of the neutron star is left on 
an eccentric and slightly wider orbit with the neutron
star no longer filling its Roche lobe. The behaviour of the 
separation as a function of time shown in Fig. 1 is very similar
to that seen in hydrodynamical numerical simulations (eg Lee 2000;
Rosswog et al 2004). 
One limitation of essentially all hydrodynamical numerical simulations
of BH--NS binaries is that they are only able to integrate for up to
0.1 s or so after the first phase of mass transfer. With our
much simpler two--body code, we are able to follow the evolution
of the BH--NS system through subsequent phases of mass transfer. 
\begin{figure}
\resizebox{8truecm}{!}{\includegraphics{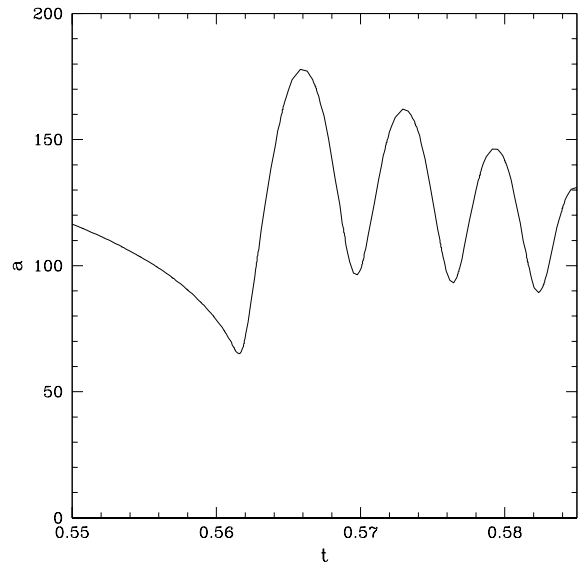}}
\caption{Separation (in km) as a function of time (in s) between
the black hole ($M_{\rm bh} = 14 M_\odot$) and neutron star. 
A short burst of mass transfer occurs at $t \simeq 0.56$ leaving
the binary somewhat eccentric and detached.}
\end{figure}
\begin{figure}
\resizebox{8truecm}{!}{\includegraphics{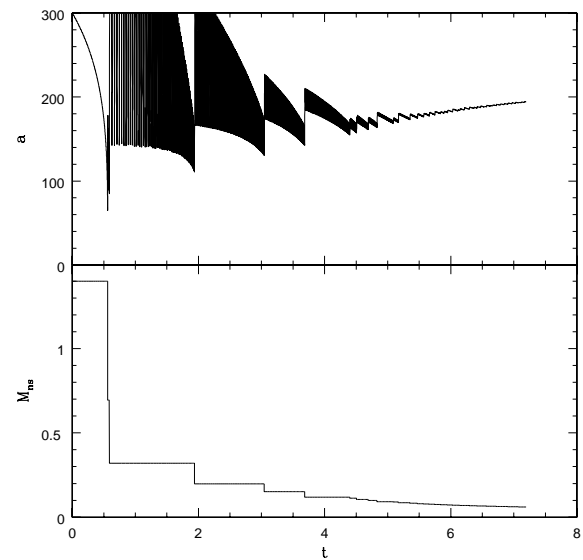}}
\caption{Top figure: separation (in km) as a function of time (in s) between
the black hole ($M_{\rm bh} = 14 M_\odot$) and neutron star. Bottom
figure: neutron--star mass (in $M_\odot$) as a function of time (in s).}
\end{figure}
This is shown in Fig. 2, where we plot the separation as a function
of time in the top half of the figure, and the neutron--star mass
as a function of time in the lower half of the figure.
From this figure we see that the neutron star quickly loses most
of its mass. The mass transfer occuring at about 0.5 s and 2 s
gives the neutron star a relatively large kick, placing it into
very eccentric orbits and increasing the pericentric separation
to over 150 km.
Once the mass falls to about 0.2 M$_\odot$ the amount
of mass transferred declines (because of a lower inspiral rate, 
$\dot{a}$, when the system comes into contact). The separation also
shows very little increase when mass transfer occurs  at this point.
Also at such low masses, the neutron star radius increases rapidly
as the mass approaches the minimum mass (e.g. Figure 1, Davies et al. 2002).
We find that, after allowing for an increase in the
neutron star radius once the mass drops below 
0.2 M$_\odot$, 
a brief phase of mass transfer no longer detaches the system, i.e. 
mass transfer continues as the NS is still filling its Roche lobe.
Rather than having a disrete burst of mass flow onto the black hole,
the neutron star will experience continuous mass loss once its
mass drops to below about 0.2 M$_\odot$. We find that this
result is true for all black--hole  masses we considered.
($3.5 M_{\odot} \leq M_{\rm bh} \leq 14.0 M_{\odot}$).  

In the results shown earlier,
we assumed that half of the angular momentum was returned
to the donor. This is a reasonable figure; we would expect
some angular momentum to be returned to the donor although
some might well be carried into the black hole with the gas.
Also, our results shown earlier are in good overall agreement
with hydrodynamical simulations
(Lee 2000; Ruffert \& Janka 2003, Rosswog et al 2004).
We see similar amounts of mass transfer from the neutron
star to the black hole during the first phase of mass transfer
and what remains of the donor is left on a similar orbit in our
calculations and hydrodynamical simulations.
Changing the fraction of angular momentum returned to the donor
does affect the subsequent trajectory of the donor;
if more angular momentum is returned to the donor, it will be left
on a wider orbit. However the system will still spiral in to contact
again, hence
the overall picture of the evolution of the system remains unchanged.
The neutron star mass is reduced to 0.2 M$_\odot$
in discrete phases of mass transfer on a timescale of $\sim 1-10$ s.

\section{Evolution at the minimum mass}
To investigate what is likely to
happen next, we must consider the response
of the neutron star to mass loss and compare this to the response
of the Roche lobe.
As given in equation (7), the Roche lobe radius can be written
as

\begin{equation}
R_L \propto a {\left( M_{\rm ns} \over M \right)}^{1/3}
\end{equation}

Differentiating this with respect to time results in

\begin{equation}
{\dot{R}_L \over R_L} = {\dot{a} \over a} + {1 \over 3}{\dot{M}_{\rm ns} 
\over M_{\rm ns}}
\end{equation}

For what follows we assume conservative mass transfer (i.e.
$M_{\rm bh} + M_{\rm ns} = $ constant: all of the mass lost by the NS
is accreted by the BH). While in practice this may not be the case
any deviations away from conservative mass transfer act to
destabilise the mass transfer further and encourage the
rapid formation of a disc.

The total angular momentum in the system is given by
\begin{equation}
J = M_{\rm bh} M_{\rm ns} {\left( Ga \over M \right)^{1/2}},
\end{equation}
which can be differentiated to give,
\begin{equation}
{\dot{J} \over J} = {\dot{M}_{\rm bh} \over M_{\rm bh}} + 
{\dot{M}_{\rm ns} \over M_{\rm ns}} + {\dot{a} \over 2a}\\
\end{equation}
Thus combining equations (11) and (13) gives us,

\begin{equation}
{\dot{R}_L \over R_L} = 2 {\dot{J} \over J} -  2 {\dot{M}_{\rm ns}
\over M_{\rm ns}} 
{\left[ {5 \over 6} - {M_{\rm ns} \over M_{\rm bh}} \right]}
\end{equation}
Including the response of the neutron star as it loses mass,
and assuming the neutron star continues to just fill its Roche lobe
as mass is transferred, we  arrive at the following equation
\begin{equation}
{\dot{M}_{\rm ns} \over M_{\rm ns}} = {{\dot{J} / J} \over \left( 
{{5 \over 6} + {\zeta
\over 2} - {M_{\rm ns} \over M_{\rm bh}}} \right)}
\end{equation}
where $\zeta$ relates the change in the neutron--star radius
to the change in its mass and is given by
\begin{equation}
{\dot{R}_{\rm ns} \over R_{\rm ns}} = \zeta \times { \dot{M}_{\rm ns}
 \over M_{\rm ns}} 
\end{equation}
For a given amount of angular momentum loss $\dot{J}$, equation (15)
gives the required value of $\dot{M}_{\rm ns}$ for the system to transfer
material in a stable fashion (i.e. where the neutron star just fills its
Roche lobe).  A key feature of equation (15) is that as $\zeta$
decreases towards $-5/3 - 2 M_{\rm ns}/M_{\rm bh}$, the required mass transfer
rate becomes increasingly large, to the point where the neutron star
transfers all its mass to the black hole in less than one orbital
period. In other words, {\em the neutron star gets shredded}. The 
exact neutron--star mass at which this occurs depends on the mass--radius
relation  but must be at a value slightly lower than 0.2 M$_\odot$
where the radius begins to expand rapidly (eg see Figure 1 in 
Davies et al 2002).

Rosswog, Speith and Wynn (2004) calculate the conditions under which
the flow from a neutron star form an accretion disc (rather than
directly accreting on to the BH). They find that for co--rotating systems
and low mass ratios ($M_{bh} / M_{ns} < 5$) disc formation is unlikely.
However in our scenario the final mass of the neutron star is
almost always $< 0.2$M$_{\odot}$, so the mass ratio
is almost always $\gg 5$. Under these circumstances it is inevitable that
the circularisation radius is greater than both the Schwarschild radius and
the innermost stable orbit.

We conclude that in BH--NS systems where mass transfer is
initially episodic, the neutron star is shredded once its
mass is reduced to about 0.15 -- 0.20 M$_\odot$, producing a disc
containing a large fraction of the material (some mass loss may occur)
with a radius in the range 150 -- 250 km (depending on the black--hole
mass). The mass within this disc is sufficient for GRB formation. 
We note that the energy released by the formation of
nuclei (of order 10 MeV/nucleon) is insufficient to eject most
of the material from the system provided $M_{\rm bh} \ge 3 M_\odot $.

An interesting consequence of the evolution described above is that
the mass of the final, GRB--producing disc is largely independent
of the initial mass of the NS (or BH). Since formation occurs at
some minimum NS mass the majority of this mass is used in
the formation of the disc and be approximately the same 
for each system. Figure 2 shows that the fractional loss of mass
per encounter is sufficiently small (for the later mass transfer events which
occur closer to the minimum mass) that the range of NS mass on the
contact at which disc formation occurs is low. The disc
in BH--NS mergers could thus offer a standard energy reservoir, making the
events standard candles.

\section{Conclusions}
We have considered the final merger of BH--NS systems.
The entire process, beginning with the first phase of mass transfer
and ending with the disruption of the neutron star, 
has a relatively long duration 
(a few seconds, or equivalently many thousands of orbital periods).
Complete hydrodynamical modelling of the entire merger process
is therefore impossible. 
Instead we have used a simpler, semi--analytical treatment for the 
episodes of mass transfer from the neutron star to the black hole.
In our model we evolve the BH and NS as point masses whose orbits 
circularise and shrink via gravitational radiation losses. 
This drag force brings the NS and BH into contact as the NS fills
its Roche lobe. We assume instantaneous mass transfer from the
NS to the BH, with some
fraction of angular momentum being transferred back to the NS on an
orbital timescale. This places what remains of the neutron star
on an eccentric and somewhat wider orbit with the neutron
star no longer filling its Roche lobe. The system then
spirals together via the emission of gravitational radiation
until the neutron star fills its Roche lobe for a second time.
More material is transferred and the neutron star receives
a second kick. This process repeats until the neutron star
reaches a mass slightly below 0.2 M$_\odot$ when its radius
increases rapidly on mass loss. At this point, the neutron star
remains in contact with the black hole and  a (brief) continuous
phase of mass transfer probably ensues. 
The mass--radius relation for the neutron star is
so steep at this point that the mass transfer is
dynamically unstable and  the neutron
star is shredded and forms a disc (with a radius
in the range 150 -- 250 km) around the black hole. 
This disc is sufficiently massive (ie around $0.1 M_{\odot}$) for
GRB production and therefore the prospect that 
BH--NS systems can produce observable GRBs is good. Furthermore
the independence of disc mass on the intial NS mass indicates
that GRBs produced via this method may have an approximately
standard energy.

\section*{Acknowledgments}

MBD acknowledges the support of a research fellowship awarded by 
the Swedish Royal Academy of Sciences. ARK acknowledges support from PPARC and
a Royal Society Wolfson Research Merit Award. AJL acknowledges receipt
of a PPARC studentship. We acknowledge the benifits of collaboration
within the European Union Research and Training Network "Gamma-Ray Bursts:
An Enigma and a Tool".

\label{lastpage}

{}

\end{document}